# CONGESTION PRICING IN A WORLD OF SELF-DRIVING VEHICLES: AN ANALYSIS OF DIFFERENT STRATEGIES IN ALTERNATIVE FUTURE SCENARIOS

Michele D. Simoni[a], Kara M. Kockelman[b], Krishna M. Gurumurthy[c], Joschka Bischoff[d]

[a] Dept. of Civil, Architectural, and Environmental Engineering, University of Texas at Austin, m.simoni@utexas.edu

[b] Dept. of Civil, Architectural, and Environmental Engineering, University of Texas at Austin, kkockelm@mail.utexas.edu

[c] Dept. of Civil, Architectural, and Environmental Engineering, University of Texas at Austin, gkmurthy10@utexas.edu

[d] Department of Transport Systems Planning and Transport Telematics, Technical University of Berlin, bischoff@vsp.tu-berlin.de

## ABSTRACT

The introduction of autonomous (self-driving) and shared autonomous vehicles (AVs and SAVs) will affect travel destinations and distances, mode choice, and congestion. From a traffic perspective, although some congestion reduction may be achieved (thanks to fewer crashes and tighter headways), car-trip frequencies and vehicle miles traveled (VMT) are likely to rise significantly, reducing the benefits of driverless vehicles. Congestion pricing (CP) and road tolls are key tools for moderating demand and incentivizing more socially and environmentally optimal travel choices.

This work develops multiple CP and tolling strategies in alternative future scenarios, and investigates their effects on the Austin, Texas network conditions and traveler welfare, using the agent-based simulation model MATSim. Results suggest that, while all pricing strategies reduce congestion, their social welfare impacts differ in meaningful ways. More complex and advanced strategies perform better in terms of traffic conditions and traveler welfare, depending on the development of the mobility landscape of autonomous driving. The possibility to refund users by reinvesting toll revenues as traveler budgets plays a salient role in the overall efficiency of each CP strategy as well as in the public acceptability.

## INTRODUCTION

Recent advances in autonomous vehicles (AVs) are generating much discussion across academia, industry, and news media. Considerable progress has been made in AV technologies, thanks to investments by technology companies and auto manufacturers (Muoio, 2017), along with support of public institutions (Kang, 2016b).

Since driverless passenger vehicles represent a new travel option, many passenger-trips made via existing, traditional modes, like cars, public transit, and bikes, will be replaced by trips made using AVs and SAVs. While AV and SAV benefits may accrue in improved accessibility, road safety and energy consumption (Fagnant and Kockelman, 2015), network congestion impacts are less well understood and may be quite problematic (Litman, 2017; Wadud et al., 2016).

On one hand, automated technologies can ultimately improve network performance by reducing traffic crashes (and the delays they entail) and eventually increasing traffic throughput by ensuring tighter headways between vehicles and by making better use of intersections. However, AVs and SAVs will likely increase the number and the distance of motorized trips by eliminating the burden of driving and by making car-travel more accessible (to persons with disabilities and those not owning cars, for example). Congestion may dramatically worsen, and demand management options will become even more valuable along congested corridors and in urban regions.

Charging drivers for the delays or congestion they cause (to those behind them, for example) is a well-known concept among economists, traffic engineers and transport professionals. Various congestion pricing policies now exist in cities like Singapore, London (UK), Stockholm (Sweden), Milan (Italy), and Gothenburg (Sweden) (Litman, 2018). Most are limited to rather simplistic cordon-based or area-based tolls that do not vary by congestion. Smartphones and/or connected vehicles offer cities, states and nations an opportunity to implement more economically efficient and behaviorally effective strategies, thanks to advanced communication and location capabilities and fast information sharing.

This study investigates the effects of different congestion pricing strategies in future scenarios characterized by strong market penetration of AVs and SAVs. Strategies include a travel time-based charge that varies with the Austin, Texas region's overall network condition and a time-varying link-based tollthat reflects marginal delay costs at the link level. The traffic and social welfare impacts of these policies are investigated and compared to those of two much simpler but rather classic strategies: a distance-based toll and a flat facility-based toll (for the most congested 2 to 4% links of the network links). To reflect the technology's uncertain development costs, capabilities and adoption rates, this work estimates two distinctive technology-adoption scenarios: one with relatively high private AV reliance and the other with high SAV uptake.

Use of congestion pricing in AV and SAV scenarios is relatively unexplored, with the exception of a few theoretical studies (as described below). This paper's simulations use the multi-agent travel-choice model MATSim (www.matsim.org). MATSim enables simulation of tens of thousands of individuals and self-driving vehicles. In this specific study, travelers' behavioral responses to CP strategies include changes in departure times, routes, activity engagements and modes, while destinations are considered fixed. Although MATSim allows for detailed analyses of a wide range of road transportation externalities (such as emissions, noise and road damage), this study focuses on congestion costs.

The remainder of this paper offers the following: a brief discussion of AVs' mobility impacts and their implications for congestion pricing, a description of the agent-based modeling framework used here, discussion of several future mobility scenarios and alternative congestion pricing strategies, analysis of the mobility and system welfare effects of such strategies, conclusions and policy recommendations.

## MOBILITY IMPACTS OF SELF-DRIVING VEHICLES AND CONGESTION PRICING

Researchers' interest in AVs' impacts on travel behavior and traffic conditions is currently strong, with different studies now having investigated issues ranging from travel behavior to advanced traffic signal control applications. Milakis et al. (2017) provide a relatively comprehensive overview and are recommended for their in-depth literature review; however, the research community's major findings are summarized below, followed by a discussion of congestion pricing.

### Travel Costs and Traveler Preferences

AV travel costs will differ from those of conventional vehicles due to several factors. Fixed costs will be somewhat higher due to expensive hardware and software required (Litman, 2017). However, thanks to lower user effort and safer driving, travel time and operating costs may fall. Parking costs can also be a key consideration for many travelers (Litman and Doherty, 2011), and "parking search" time can represent a significant portion of travel time in very busy settings (Shoup, 2006).

To date, there is little consensus in the scientific community about the impacts of AVs on travelers' value of travel time (VOTT). While some argue that autonomous travel might not substantially affect travelers' overall time-cost perceptions (Cyganski et al., 2015; Lenz et al., 2016; Yap et al., 2016), most assume and/or estimate (using stated-preference data) a lower VOTT (see, e.g., van den Berg and Verhoef, 2016; Lamotte et al., 2017; Zhao and Kockelman, 2017; Bansal and Kockelman, 2017).

SAV costs are likely to be under $1/mile, thanks to no driver wages (Fagnant and Kockelman 2018; Loeb et al., 2018; Chen et al. 2016; Bosch et al., 2017). Most of the previous literature on SAVs has focused on the effects of replacing trips and on the required fleet sizes for the full replacement of conventional trips (Burns et al., 2013; Spieser et al., 2014; Fagnant et al., 2015; Chen et al., 2016; Bischoff and Maciejewski, 2016; Gurumurthy and Kockelman, 2018). Only a few studies have focused on investigating their adoption from a travel demand modeling perspective (Correia and van Arem, 2016; Krueger et al., 2016; Haboucha et al., 2017; Scheltes and Correia, 2017; Martinez and Viegas, 2017).

## Changes in traffic conditions

Autonomous driving will affect traffic conditions in several ways. Thanks to reduced reaction times and shorter following distances, road and intersection capacity may eventually increase (Dresner and Stone, 2008), thereby reducing delays. Cooperative technologies like vehicle-to-vehicle (V2V) and vehicle-to-infrastructure (V2I) communication can improve network performance (Tientrakool et al., 2011; Shladover et al., 2012; Hoogendoorn et al., 2014). Interestingly, Talebpour and Mahmassani (2016) use microsimulations to show how automation can play a larger role than connectivity in terms of capacity impacts. Most of these improvements will likely occur only for large adoption rates of self-driving and connected vehicles.

However, reduced driving burdens, lower travel cost, and improved transport access will probably increase car use and travel distances in the near term, delivering more congested traffic conditions long before connected and autonomous vehicle (CAV) technologies can resolve many capacity issues (Gucwa, 2014; Fagnant and Kockelman, 2015; Milakis et al., 2017). Empty SAVs driving in between trips may compound this effect, with actual increases depending on each region's policies, mode options, parking costs, and trip-making patterns, for example. For Berlin, Germany, SAVs (without dynamic ride-sharing among strangers) added 13% VMT (Bischoff and Maciejewski, 2016). Re-balancing vehicles during off-peak times may help to minimize the effect of those trips (Winter et al., 2017; Hörl et al., 2017). However, the possibility of pooling multiple travelers on SAVs with dynamic ride-sharing could reduce the overall VMT (Fagnant and Kockelman, 2018).

## Congestion pricing for AVs

The idea of charging individual travelers for the marginal external congestion cost of their trips (associated with the delays caused to other users) has a long tradition in the transportation economics and engineering field (Pigou, 1920; Walters, 1961; Vickrey, 1963, 1969; Beckman, 1965). During the past fifty years, several studies have investigated the topic of congestion pricing to identify optimal strategies and to add more realistic features to their models based on travelers' behavior and infrastructure performance (Small and Verhoef, 2007; de Palma and Lindsey, 2011). The adoption of AVs and the emergence of mixed-flow scenarios bring new theoretical and practical challenges for pricing.

From a theoretical perspective, AVs will affect the demand curve (volumes of road users). As discussed in the following section, the possibility to perform other activities while driving and the different level of comfort, will likely affect AV users' travel costs and perceived VOTT. As van den Berg and Verhoef (2016) highlight, a decrease of VOTT would correspond to a reduction in queuing costs for AV users, ultimately increasing their acceptance of increased travel time. Hence, from a marginal cost perspective, this phenomenon could hurt conventional users, whose travel costs have not changed.

In addition to that, AVs will affect the supply curve (response of the transportation system to the demand). As discussed later in the paper, AV implementation will likely yield improvements in network performance. Lamotte et al. (2017) considers some of these aspects by investigating the problem of optimal allocation of road infrastructure (between conventional and automated cars) and tolls with a bottleneck model. In case of shared road infrastructure, congestion phenomena become more complicated. From a traffic flow perspective, increasing levels of AVs would delay the onset of congestion (shift of Critical Density in the Fundamental Diagram[1]), allowing additional demand without compromising traffic conditions. However, larger portions of AVs will also accelerate the deterioration of traffic conditions in the congested traffic regime (increase of congested wave speed). Hence, from a marginal cost perspective, higher shares of AVs would be beneficial to all travelers up to a certain level of traffic demand (capacity).

From a practical standpoint, the high level of information and communication characterizing autonomous driving might favor the introduction of more advanced tolling strategies. Ideally, tolls should reflect changes in travel costs depending on category of road user, time of the day, real-time traffic conditions, trip purpose, and presence of transportation alternatives like public transit (Vickrey, 1997; Arnott, 1998). However, current road and congestion pricing strategies typically include: facility-based tolls (on bridges, tunnel, highways); cordon-based tolls (applied when entering an area, like in Milan, or when crossing the cordon in either direction like in Stockholm and Gothenburg); area-based tolls (applied when driving inside the area, like in London); and distance-based tolls (like in heavy goods vehicles charges in Germany).

In the past, research in the field has been characterized by a clear distinction between "first-best pricing" solutions, with strong analytical frameworks and the absence of constraints, and "second-best pricing" solutions, characterized by sub-optimal, but more feasible mechanisms (Verhoef, 2002; de Palma et al., 2004; Zhang and Yang, 2004; May et al., 2008; Lawphongpanich and Yin, 2010).

The possibility of exchanging traffic and charges information in real-time to all the connected vehicles would allow for more advanced pricing strategies that vary in time, space and level of charge more dynamically. In this sense, AV technologies could almost bring second-best pricing systems to first-best pricing ones (as far as congestion externalities are concerned). Driverless technologies can also facilitate congestion pricing in other ways. First, tolling systems could become more feasible thanks to advanced communication technologies (wireless, GPS), cheaper than the current tolling systems based on dedicated short-range communications (DSRC) and automated license plate recognition since they do not need additional road infrastructure. Second, the fact that "smart" AVs can compute and communicate tolls and routing options to travelers would help to keep pricing schemes understandable and transparent. This might eventually increase public acceptability of congestion pricing (Gu et al., 2018).

In the "Congestion Pricing Schemes" section, we present two congestion pricing schemes that leverage the advanced communication and computation capabilities of AV-SAVs to derive schemes closer to the concept of first-best pricing. To the best of our knowledge, congestion pricing schemes involving both privately owned AVs and SAVs are still relatively unexplored except for a few studies in specific simulation environments (agents' route choice under dynamic link tolls) (Sharon et al., 2017) or involving broader external costs for some specific modes (limited to SAVs and conventional cars) (Kaddoura et al., 2018). In this study, we provide a comprehensive analysis of different potential future scenarios (accounting for different market penetrations of AVs and SAVs) and alternative congestion pricing schemes.

## MODELING AVs AND SAVs WITH AN AGENT-BASED MODEL

In this section after providing a brief overview of the agent-based model MATSim, we present a description of our modeling framework. We then focus on the modeling of AVs and SAVs.

### General Framework of MATSim

MATSim simulates the daily plan-set of all agents and considers endogenous mode choice, departure time choice and route choice, making it a fully dynamic model. As opposed to models that use single trips, this model allows for predictions on reactions to demand management strategies, such as tolls during the span of a day, thus accounting for a higher level of realism. In fact, trips are typically linked to each other as a part of a daily plan and are not that meaningful as stand-alone trips (Balmer et al., 2006). Activities often have higher importance in the daily schedule than trips that simply represent connections among them. Since MATSim represents traffic behavior at a highly disaggregated level by modeling individual agents (with different socio-demographic characteristics), it is possible to investigate the effects of transport policies on travel behavior and traffic in more detail than in traditional 4-steps models (Kickhöfer et al., 2011). The overall process (Figure 1) can be summarized in the following stages:

- Each agent independently develops a plan that expresses its preferences in terms of activities, trips and their schedules during the day (Initial Demand).

- The agents simultaneously perform all the plans in the mobility simulation's (Mobsim) physical system. Congestion phenomena are modeled using a queue model, which takes both the physical storage capacity and the actual throughput (flow capacity) of a link into account.

- To compare the performance of different plans, each plan is scored using a utility-type function (Scoring).

- Agents are able to remember their plans and improve them during the simulation by means of a learning algorithm (Replanning). During implementation the system iterates between plan generation and traffic flow simulation.

- The cycle is run until the system has reached an equilibrium where no agent can improve its score anymore (Analyses).

---

[1] The Fundamental Diagram relates a roadway's traffic flow and density values (Lighthill and Whitman, 1955; Richards, 1956).

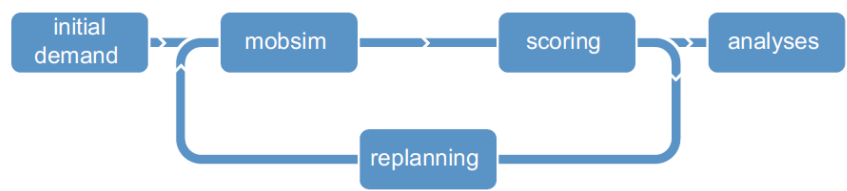

*Figure 1: MATSim cycle (source: Horni et al., 2016)*

The choice model generally adopted in MATSim is equivalent to a multinomial logit model. Since the amount of plans in the memory of agents is limited, the worst performing one is replaced by a new one at each iteration. Thanks to this feedback mechanism, agents are able to improve their plans over several iterations until the system reaches the "relaxed" state when agents cannot significantly improve their plans and the outcome of the system becomes stable. This state is also referred to as the agent-based stochastic user equilibrium (Nagel and Flotterod, 2009).

For further information about the simulation framework MATSim, see Horni et al. (2016).

**Choice dimensions and parameters**

In this study, daily itineraries or agents' plans contain up to five different activity types: "Home," "Education," "Work," "Shopping," and "Leisure," which can be linked via several possible trip-chain combinations. As shown in Figure 2, each plan describes a tentative schedule of activities (with their locations) and travel choices to reach them. Plans, which reflect typical activity schedules, have been derived based on real travel demand data (Liu et al., 2017). For example, at 7 AM, "Home" represents the majority of the activities performed (around 90%), while at 10 AM, agents are involved more evenly with different activities ("Home":25%, "Education":10%, "Work": 27%, "Shopping":21%, and "Leisure":17%). Plans can be improved by changing the time of departure, varying the route and/or choosing a different transport mode through a series of choice modules. Agents' travel choices are modeled in MATSim through an iterative learning mechanism based on a quantitative score, referred to as utility (Eqn. 1). In each iteration, agents choose from an existing set of daily plans according to a multinomial logit model.

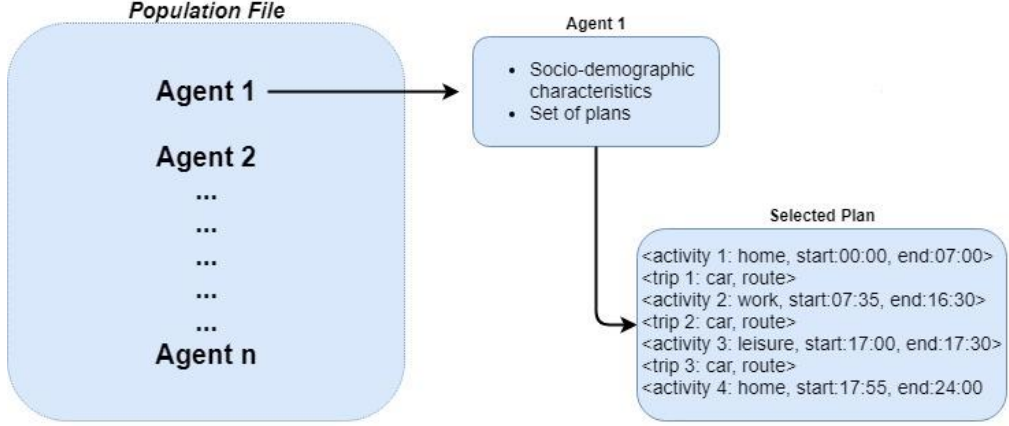

*Figure 2: Example of two agents' plans*

Every daily plan is associated with a utility score by taking into account the trip-based travel disutility and utility from performing an activity:

$$V_{plan} = \sum_{i=1}^{n} (V_{act,i} + V_{trip,i})$$ (1)

where $V_{plan}$ is the total utility of a daily plan; $n$ is the total number of activities or trips; $V_{act,i}$ is the utility for performing activity $i$; and $V_{trip,i}$ is the utility of the trip to activity $i$. The first and the last activity are wrapped around the day, and handled as one activity. Thus, the number of activities and trips is the same. Each mode's trip-related utility is calculated as follows:

$$V_{q,i} = \beta_{0,q} + \beta_{t,q} \cdot t_{i,q} + \beta_c \cdot c_{i,q}$$ (2)

where $\beta_{0,q}$ corresponds to the alternative specific constant (ASC) of mode $q$; $t_{i,q}$ corresponds to the travel time of leg $i$ traveled with mode $q$; $\beta_{t,q}$ corresponds to the marginal utility of traveling by mode $q$; $c_{i,q}$ corresponds to the monetary cost of leg $i$ traveled with mode $q$; and $\beta_c$ corresponds to the marginal utility of monetary cost.

To calculate the positive utility gained by performing an activity, a logarithmic form is applied (Charypar and Nagel, 2005; Kickhofer et al., 2011):

$$V_{act,i}(t_{act,i}) = \beta_{act} \cdot t_i^* \cdot \ln\left(\frac{t_{act,i}}{t_{0,i}}\right)$$ (3)

where $t_{act}$ is the actual duration of performing an activity, $t_i^*$ is an activity's 'typical' duration, and $\beta_{act}$ is the marginal utility of performing an activity at its typical duration. In the equilibrium, all activities at their typical duration are required to have the same marginal utility; therefore, $\beta_{act}$ applies to all activities. $t_{0,i}$ is a scaling parameter linked to an activity's priority and minimum duration. In this study, $t_{0,i}$ is not relevant, since activities cannot be dropped from daily plans.

The value of travel time saving (VTTS) is derived as follows:

$$VTTS = \frac{\beta_{act} - \beta_{t,q}}{\beta_c}$$ (4)

where $\beta_c$ corresponds to the marginal utility of money.

The travel options modeled in this study include: car, public transit, bike and walk (modeled jointly), AV, and SAV. The behavioral parameters for car and public transit used in this study are based on Tirachini et al. (2014) and Kaddoura et al. (2015) and have been adjusted to reflect the current travel costs in the U.S. (2017). The parameters used for the simulation are summarized in Table 1. Since the simulation approach does not explicitly account for parking costs and walking times of car users, we have derived an alternative specific constant $\beta_{0,car} = -0.1$. This value roughly corresponds to $0.30 per trip and accounts for the fact that current parking charges applies to just the most central portion of the network (downtown). In addition, car users pay a monetary cost proportional to the distance traveled corresponding to $0.30 per mile. Since waiting, egress and access times are not modeled in these experiments, public transit (PT) has been recalibrated, yielding an alternative specific constant $\beta_{0,PT} = -1.5$. This value also accounts for the average ticket cost and for Americans' and Austinites' reluctance in using public transit. In a similar fashion, the alternative specific constant for walking/biking has been set to $\beta_{0,active} = -0.2$. Similar to Kaddoura et al. (2015), the marginal utility of traveling by car is set to zero. Even if this value is set to zero, traveling by car will be implicitly punished by the opportunity cost of time (Horni et al., 2016). In this study, the marginal utility of money $\beta_c$ is equal to 0.79 such that the VTTS for car users corresponds to about $18 per hour. This value has been obtained according to the recommendations from the USDOT (2011).

The AV parameters have been largely derived from Kockelman et al.'s (2017) and Bosch et al.'s (2017) work. A privately owned and operated AV is assumed to cost $0.20 per mile since fuel economy, insurance costs, and maintenance costs should be lower than those of a conventional car (Bosch et al., 2017). We assume AVs to have a null alternative specific constant in order to account for parking and walking time reductions. The marginal disutility of traveling is set equal to +0.48 to reflect a marginal cost of traveling equal to 50% of those of car users (corresponding to a VTTS of about $9 per hour), in line with Gucwa (2014) and Kim et al. (2015).[2]

As for SAVs, we assume the same alternative specific constant and marginal cost of traveling of AVs since they are used by only one individual or party at a time. Unlike AVs, SAVs are characterized by waiting times depending on the availability of vehicles. We assume the monetary costs to be composed of a fixed flat fee, and variable distance fare and time fare, depending on the scenario (see the following sections for further details).

*Table 1: Mode choice parameters used*

| Travel Mode | $\beta_0$ | $\beta_t$ |
|---|---|---|
| Car | -0.1 | 0.00 |
| Public Transit | -1.5 | -0.36 |
| Walk/Bike | -0.2 | 0.00 |
| AV | 0.0 | +0.48 |
| SAV | 0.0 | +0.48 |

In addition to travel choices, agents can modify the start time and duration of each activity in their plan-set to reflect aspects like the optimal/target duration for the activity type, and site opening and closing times (Table 2). Activities performed outside open/feasible times do not offer any added utility. Furthermore, agents are subject to schedule penalty costs for being early or late according to Vickrey's parameters: *α, β,* and *γ* (Arnott et al., 1990). Although agents' decision to drop activities is not explicitly modeled, when transportation costs are very high, agents' could extend their activities and render participation in following activities impossible.

---

[2] Note that, in MATSim, setting a positive marginal disutility of traveling does not imply a gain of score from the trip since agents are punished by the opportunity cost of time (loss from not being able to perform the desired activity)

*Table 2: Travelers' out-of-home activity attributes*

| Activity Type | Optimal duration | Opening time | Closing time |
|:---:|:---:|:---:|:---:|
| Home | 14 | Undefined | Undefined |
| Education | 5 | 08:00 | 22:00 |
| Work | 7 | 07:00 | Undefined |
| Shopping | 1 | 09:00 | 01:00 |
| Leisure | 2 | 09:00 | 01:00 |

**Simulation of shared mobility services and road capacity increase**

The simulation of SAVs is performed by means of an extension to MATSim that allows solving dynamic vehicle routing problems (DVRP) using a dedicated module (Maciejewski et al., 2017). The DVRP contribution reproduces dynamically demand-responsive modes such as conventional taxis and ride hailing services. As opposed to the standard vehicle routing in MATSim, which is conducted before each iteration starts, the DVRP module allows an online dispatch of SAVs. Vehicle dispatch is generally started the moment an agent wishes to depart using such a mode (and SAVs cannot be booked in advance here).

For the simulation of large fleets of SAVs, a straightforward, rule-based dispatch algorithm is used, which has been applied in previous case studies with more than 100,000 vehicles (Bischoff and Maciejewski, 2016). The algorithm aims at reducing the required number of vehicles during peak hours and thus minimizes the required fleet to serve all the received requests. This issue is addressed by a "demand-supply balancing" vehicle dispatch strategy, in which the system is classified into two mutually-excluding categories, namely oversupply, with at least one idle SAV and no open requests, and undersupply, with no idle SAVs and at least one open request. Both states are being handled in different ways. In the first case, when a new request is placed, the nearest vehicle is dispatched towards it. In the latter case, when a vehicle becomes idle it is dispatched to the nearest open request. In times of oversupply, requests are served immediately, whereas in times of undersupply, a vehicle will first be dispatched to requests waiting in close proximity and thus may leave requests waiting for a longer time. This helps to maximize the throughput of the system. Despite its simplicity, this strategy provides solutions that are close to those of more complex methods, such as solving iteratively the taxi assignment problem (Maciejewski et al. 2016).

In order to account for the capacity increase resulting from reduced reaction times and shorter following distances, a specific MATSim module is adopted that allows for traffic simulation of mixed autonomous/conventional flows (Maciejewski and Bischoff, 2017). This is achieved by lowering the capacity (maximum flow) required by AVs to travel on a link by a factor of 1.5. This means that a link that may otherwise be passed by a maximum of 1000 conventional vehicles per hour could be passed by 1500 AVs per hour. In case of mixed flows of AVs and conventional vehicles on the link, the maximum flow lies between these values, depending on the actual vehicles' mix (Figure 3) and following a flow capacity increase ratio of: $1/(1 - s + s \cdot c)$, where $s$ and $c$ represent the share of AVs and the capacity increase parameter (equal to 0.666), respectively. Hence, the benefits of increased levels of AVs traffic are not linear. The results of the model are in line with those in Levin and Boyles (2016), who proposed a multiclass cell transmission model for shared human and AV roads.

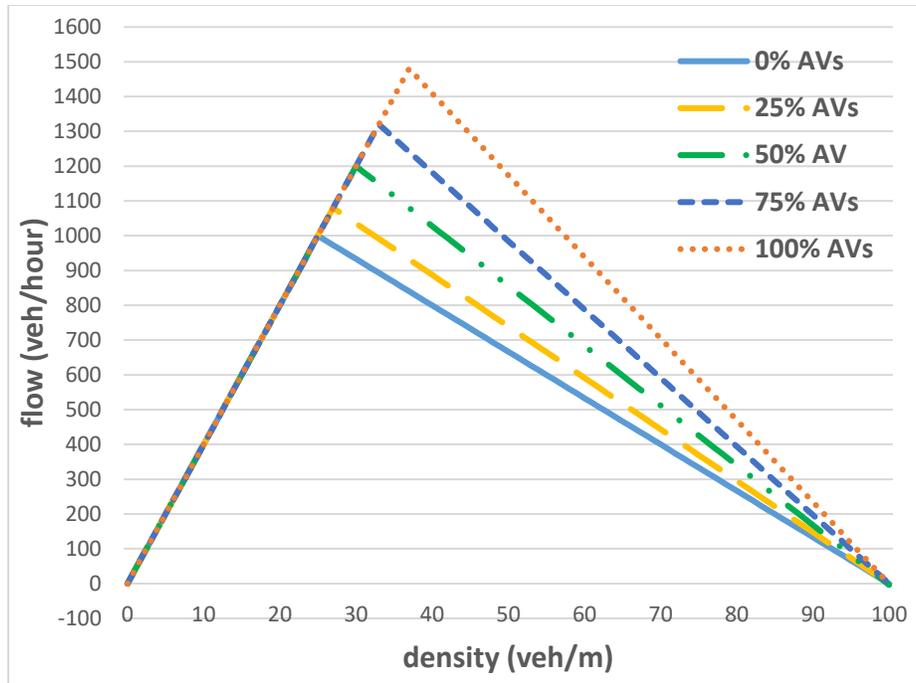

*Figure 3: Impact of different shares of AVs on traffic flow*

## SIMULATION SCENARIOS

The impacts of different pricing schemes are investigated for three different scenarios. The "Base Scenario" corresponds to a realistic simulation of the city of Austin and surroundings (Figure 4), comprising a considerable portion of the Austin metropolitan area (Greater Austin). The studied region, which includes satellite cities such as Round Rock, Cedar Park and Pflugerville, accounts in total for a population over 2 million (U.S. Census Bureau, 2017). The presence of a major trade corridor (Interstate Highway 35, through the region's heart), absence of high-quality public transit options (a single light-rail line and limited-frequency buses serve a subset of neighborhoods), and annual population-growth rates around 3.0% are some of the reasons behind the region's serious congestion issues. TomTom (2016) data rate Austin as US's 15th most congested city, with a congestion score or (average) extra travel time of 25%. The simulation's high-resolution navigation network includes 148,343 road segments (links). The population and agents' travel plans (activity chains) were obtained by adjusting Liu et al.'s (2017) year-2020 household data (based on the metropolitan transportation agency's 2020 trip tables and demographic data). Although the plans have not been formally validated, they have been adjusted to achieve realistic modal share, trip distances and durations. More than 100 types of trip-chain profiles deliver 3.5 trips per traveler per day. Each traveler (or active agent for that day) needs to travel at least once to execute his or her plans. Instead of simulating the full population, a sample of 5% (equivalent to 45,000 agents) is used here. A simulation of 150 iterations of such sample would still require between 12 and 20 hours on a super-computer. Link capacities are downsized to match the sample size. The available transportation modes (for regular, passenger travel) in the Base Case are conventional cars/passenger vehicles, public transit and walk/bike (modeled jointly). In order to reflect current trends in availability of car as a travel option, we assume 90 percent of agents have access to a car (either as a driver or passenger). In the simulations, public transit is assumed available to any traveler, although in some of the most peripheral areas, access and waiting times might be very poor.

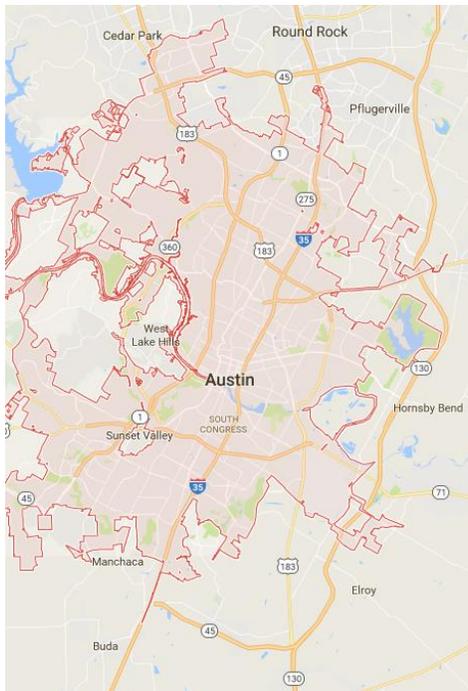

*Figure 4: Simulation Network (source: Google Maps)*

The two additional scenarios correspond to possible future scenarios characterized by the presence of AVs and SAVs. Currently, it is not clear whether AVs will mainly replace privately owned vehicles or if they are going to be adopted as shared taxis. On one hand, the auto industry is moving quickly to provide the first "partially autonomous" models (Level 3) by 2020 and full autonomous models by 2030 (Level 4 and Level 5) (Kockelman et al., 2017). Conversely, ride-sharing companies (Uber, Lyft, Didi) are already running tests (Kang, 2016a; Hawkins, 2017), making considerable investments (Buhr, 2017), and developing important partnerships (Russell, 2017) to put driverless fleets on the road within a few years. Hence, an "AV-oriented" Scenario and a "SAV-oriented" Scenario are included, to represent these distinctive trends. In the AV-oriented scenario, it is assumed that a large portion of the population will switch from car to AV (90% of agents having accessibility to car in the Base Scenario). In this scenario, the cost of AVs is lower than car cost ($0.20 per mile). SAVs are available too, but the fleet size is relatively small (one vehicle for every 30 agents) and they are characterized by lower prices than the current shared mobility services ($0.50 flat charge, $0.40/mile distance charge and $0.10/minute time charge). For example, a trip of 5 miles, from the northern suburbs to downtown, would vary approximately between $3.70 and $5.20 depending on traffic conditions. In the SAV-oriented scenario, SAVs are largely available (one vehicle for every 10 agents), whereas most of the population is still car-dependent (only 10% has access to privately owned AVs, the cost of which corresponds to $0.20 per mile). Furthermore, we assume a decrease of availability of privately owned vehicles to 60% in order to reflect a decrease of ownership (Litman, 2017). In this scenario, SAVs are characterized by lower prices than in the AV-oriented scenario (a 50% reduction), assuming that main ride-sharing companies and local authorities would stipulate agreements on prices concerning the provision of shared autonomous services. In this case, the same type of trip described above would cost approximately between $1.80 and $2.60 (slightly higher than a public transit pass).

Results of MATSim simulations in terms of modal shift are reported in Figure 5. In the Base Scenario, a car clearly appears as the dominant travel option, in line with the current situation. In the AV-Oriented Scenario and SAV-Oriented Scenario, the introduction of two additional travel options (SAVs and AVs) generates significant changes. Public transit (PT) trips decrease in the AV-Oriented Scenario (to 4% of the mode share), whereas they slightly increase in the SAV-Oriented Scenario (to 8% of the mode share). This result is partly due to the lower ownership of private vehicles, which forces a considerable portion of commuters to travel with either PT or SAVs. "Active trips" decrease to 2% and 4% respectively in the AV-Oriented Scenario and SAV-Oriented Scenario. As a result of this shift, congestion measured as daily total vehicle-miles traveled (VMT) and daily total travel delay increase in both the scenarios (Table 3). The increased capacity due to autonomous driving is offset by the increased trips particularly in the SAV-oriented Scenario by the empty SAV trips that account for 6.2% of the total VMT.

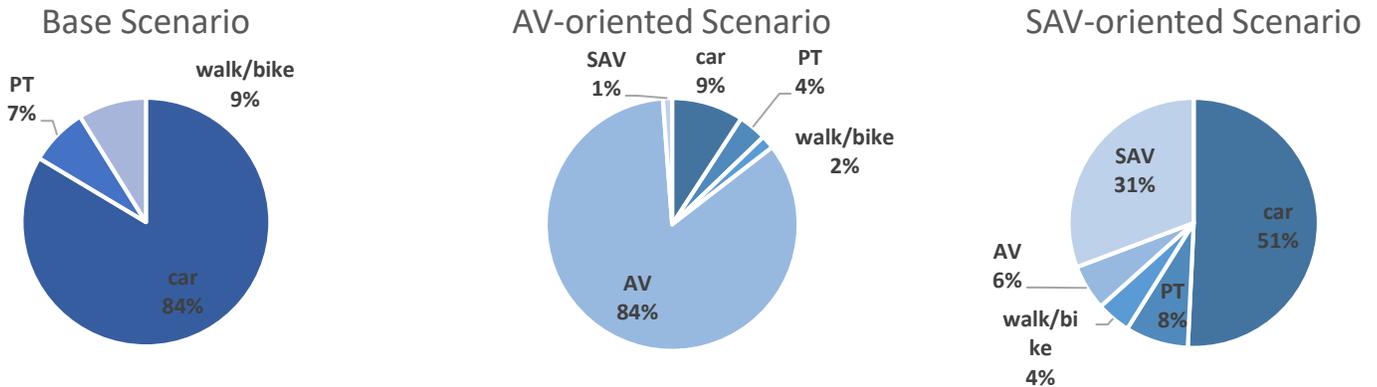

*Figure 5: Modal Split across three different scenarios*

*Table 3: Traffic conditions of the three different scenarios*

|  | **Base Scenario** | **AV-Oriented Scenario** | **SAV-Oriented Scenario** |
|---|---|---|---|
| **Total Daily VMT** | 2,671,560 mi/day | 3,104,043 | 3,271,169 |
| **VMT by Empty SAVs** | 0 | 4,741 | 201,828 |
| **Total Travel Delay** | 251,475 veh-hr/day | 405,854 | 469,123 |

## CONGESTION PRICING STRATEGIES

This study investigates the performance of four different congestion pricing strategies. A facility-based and distance-based scheme are considered "traditional schemes", since they are well known in academia and in practice. A link-based marginal-cost-pricing scheme and a travel-time-congestion dependent scheme are considered "advanced schemes" because they are more complex and require relatively new technologies (such as those of connected-automated vehicle) for optimal implementation. Because of that, the advanced schemes are assumed to be implemented only in the AV-Oriented and SAV-Oriented scenarios.

### Traditional congestion pricing strategies

Facility-based tolls are probably the most common form of congestion pricing since they do not require particularly advanced technologies for implementation. In the past, this type of scheme has been implemented mainly on tunnels, bridges and highway facilities that represent major bottlenecks. In this study, a "Link-based Scheme" is applied to the most congested links during the morning peak hours (7-9 AM) and evening peak hours (5-7 PM). The tolled links are selected based on the volume/capacity (V/C) ratio calculated on hourly basis and aggregated for the peak hour periods. A minimum threshold V/C ratio of 0.9 is chosen to identify the most congested links, resulting in the selection of about 2 to 4% of the road network (3,911 links in the Base Scenario, 4,850 links in the AV-Oriented Scenario, and 4,424 links in the SAV-Oriented Scenario). As illustrated in Figure 6, the tolled links include the most important segments of Austin's highway system, including Interstate 35 and the State Loop 1. A flat toll rate is set to all the selected links regardless of the amount of congestion and the characteristics of the link. The toll value has been derived by testing different levels of toll ranging from $0.10/link to $0.30/link in each scenario and selecting the most effective one in terms of social welfare impacts (Appendix I). The obtained values correspond to $0.1/link for the Base Scenario and SAV-Oriented Scenario, and $0.2/link for the AV-Oriented Scenario. Texan toll roads have varying charges between $0.20 and $1+ per mile, although they are limited to small portions (25 highway sections summing to less than 200 centerline-miles) with revenues mainly used for future road projects and maintenance (Formby, 2017).

Here, the Distance-based Scheme's toll varies simply with distance traveled, at a rate of $0.10 per mile (for all scenarios) between the hours of 7 AM and 8 PM. This toll was chosen to maximize agents' social welfare, as discussed above, for the Link-Based scheme (Appendix I). One could make it more time or location dependent, requiring on board GPS to keep track of each vehicle's position (and tally the owed charges before reporting back to a fixed roadside or gas-pump-side device, for example). Of course, many nations, states and regions are interested in distance-based tolls or VMT fees, especially when more fuel-efficient and electric vehicles pay relatively few gas taxes. Clements et al. (2018) discuss such tolling options, and the strengths and weaknesses of various tolling technologies.

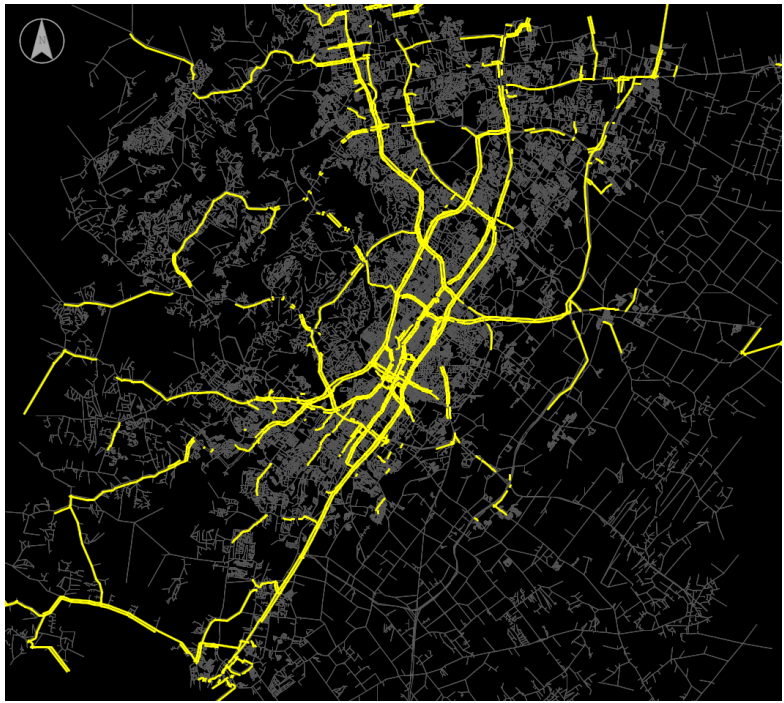

*Figure 6: Selected links in the Link-based Scheme for the Base Scenario (source VIA:Senozon)*

**Advanced congestion pricing strategies**

The first advanced congestion pricing strategy investigated consists of a dynamic marginal cost pricing (MCP) scheme at link level. In the context of road usage, MCP means charging users for the extra cost (shadow cost) that their trip causes to other travelers (Walters, 1961) due to longer travel times. According to MCP models, an optimal, static, link-based toll $\tau$ can be derived for each link such that:

$$\tau = V \cdot \frac{\partial c}{\partial V} \qquad (5)$$

where $V$ corresponds to the traffic volume on the link and $c$ corresponds to the congestion costs that can be related to $V$ by means of several functions. However, MCP presents some theoretical and practical limitations, including the dynamic nature of congestion and the difficulty of setting operationally (and socially) optimal link tolls across large networks (De Palma and Lindsey, 2011).

Communication and automation technologies installed in AV/SAVs offer the opportunity to apply different tolls on each link of a network that vary dynamically according to traffic conditions. In this proposed "MCP-based scheme," each link's cost of congestion is derived using the Fundamental Diagram (FD), which is a relation between traffic throughput (or outflow) $q$ (veh/h) and density $k$ (veh/km) (Greenshields et al., 1935). According to the FD, the throughput increases with density until reaching the critical density corresponding to the link's capacity. For values of density above the critical one, the link's throughput and (average) speed fall toward zero. Based on this concept, it is possible to estimate for each link, during a certain time interval, the amount of delay and corresponding toll such that queues can be eliminated and the traffic throughput adjusted to capacity.

Since MATSim reflects FD behavior, one can derive each link's average speed $u(k, q)$ as function of its traffic density and outflow, as follows:

$$u(k, q) = \frac{q}{k} \qquad (6)$$

Thus, for each link, the total delay accumulated during the time interval $[t, t + \Delta t]$ is:

$$d = \left[ \left( \frac{l}{u_{t+\Delta t}} - \frac{l}{u_t} \right) \cdot n \right] \qquad (7)$$

where the first term corresponds to the marginal delay per time interval, which is given by the difference of travel time on link of length $l$ at the average speed $u$ and at free-flow speed $v$, and the second term $n$ corresponds to the link users (vehicles) per time interval. The number of additional users (of the link) $\Delta n$ over the time interval (only in case of decrease of outflow and speed) can be derived as:

$$\Delta n = (q_t - q_{t+\Delta t}) \cdot \Delta t \qquad (8)$$

Hence, the marginal cost pricing charge for each link $m$, during the time interval $[t, t + \Delta t]$ can be derived as:

$$\tau_m = max \left\{ 0; \frac{d_m \cdot VTTS}{\Delta n_m} \right\} \qquad (9)$$

where VTTS corresponds to the average value of travel time. For reasons of travelers' understanding and acceptance (public acceptability), each link's charge varies over intervals of 15 minutes and it comes from aggregating traffic condition measurements across 5-minute intervals. The duration of the intervals is selected based on practical reasons. On the one hand, charges should not change too rapidly in order to allow users to adjust properly (users' adaptation). On the other hand, charges should not change too slowly in order to properly capture the dynamics of congestion. The toll has a maximum threshold value of $0.30. For practical reasons, given the size of the network, a subset of 15,020 centrally located links are analyzed here, as shown in Figure 7.

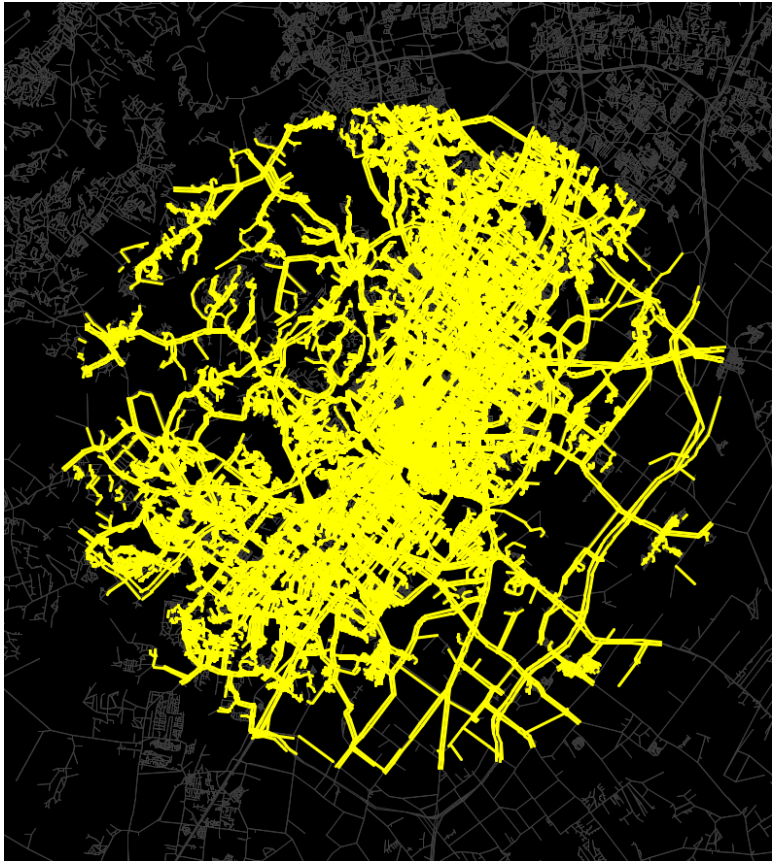

*Figure 7: Analyzed links in the MCP Scheme (source VIA:Senozon)*

The second advanced congestion pricing scheme is a joint "Travel Time-Congestion-based scheme." The main rationale behind this approach lies in the fact that simple, distance-based strategies do not reflect traffic dynamics. They can even be detrimental if drivers are incentivized to take shorter (but more congestible) routes (Liu et al., 2014). Charging users for the delay caused (at network level) during their time traveled, depending on the time of the day and on traffic conditions of the network, could obviate this problem. Hence, trips made during the more congested times will be more penalized because of longer travel times and higher tolls. Similar to transportation network companies' (TNC) surge-pricing policies, where prices vary with demand-supply ratios, in the Travel Time-Congestion-based scheme dynamic tolls are derived as follows:

$$\tau = \alpha \cdot \sigma_{[t,t+\Delta t]}(t) \qquad (10)$$

where $\alpha$ is a constant proportional parameter, which influences how quickly the theoretical optimal toll is achieved, and $\sigma$ is the network congestion dependent component. A conservative value of $\alpha = 0.1$ is assumed in both the AV and SAV-oriented scenario. The component $\sigma_{[t,t+\Delta t]}(t)$ varies every 30 minutes, based on traffic conditions measured across all six 5-min intervals in that half hour. In order to reflect changes of overall marginal cost of congestion on the network, the travel-time-congestion-dependent component is derived as follows:

$$\sigma_{[t,t+\Delta t]} = \frac{(\sum_i^M d_i) \cdot VTTS}{S \cdot r} \qquad (11)$$

where link $i$'s delay $d_i$ is calculated using Eq. 7 for the networks' $M$ links, $S$ corresponds to the total number of departures over the time period $[t, t + \Delta t]$, and $r$ corresponds to the average trip duration on the network, which is derived as follows:

$$r = \frac{L}{U} \qquad (12)$$

where $L$ and $U$ correspond to the average trip length and average free-flow speed over the network, respectively.

Since both advanced schemes seek to be consistent with traffic dynamics, which in turn depend on agents' mode, departure time and route choices, we adopt a simulation-based feedback iterative process to derive the final toll values of the vector of tolls $\bar{\tau}$ for all the links considered. Given the complexity of the problem, two stopping criteria are used here. The first one, similarly to Lin et al.'s approach (2008), uses the average difference of trip travel time (for each agent) as follows:

$$\Delta TT = \frac{1}{J} \cdot \sum_j^J \sum_i^I \frac{\left| tt_{i,j}^{k-1} - tt_{i,j}^k \right|}{tt_{i,j}^k} \cdot 100$$

where $tt_{i,j}^k$ corresponds to the travel time of agent's $j$ trip $i$ in iteration $k$. The second stopping criterion corresponds to the average change of agents' utilities, $\Delta U$. Hence, for each iteration $j$, the algorithm performs the following steps:

1. Identify toll values $\bar{\tau}_j$ for each time interval $[t, t + \Delta t]$ by means of Eq. 9 or Eq. 10.
2. Perform a MATSim simulation until new stochastic user equilibrium is reached.
3. Derive the average difference of travel time of trip $\Delta TT$ and agents' utilities $\Delta U$ between the current iteration $j$ and the previous ($j$-1).
4. Check if both meet the target value. If yes, stop. Otherwise, return to step 1.

Owing to computational limitations, only 150 iterations per simulation of MATsim could be run with the developed code. The results between the final and penultimate iterations have scores within 5% of each other, although route choices may still vary on the links used be very few agents per time interval. Even though these results are suboptimal, they can be assumed close to the final route choices.

The resulting tolls for the MCP-scheme for the AV-Oriented and SAV-Oriented Scenario are determined after 10 to 15 simulations (Figure 8). Among the 28,484 links analyzed, between 5,000 and 7,000 are tolled (across the various 15-min intervals), with an average charge of $0.02-$0.05 (per link) in each scenario.

Figure 9 illustrates tolls for the Travel Time-Congestion-based scheme for both the AV-Oriented and SAV-Oriented Scenarios. The two schemes show similar trends in the variation of the travel time toll during the peak hours, with higher charges during the morning peak. As expected, since AV travel costs are lower than SAV travel costs, the resulting levels of charge in the AV-Oriented Scenario are higher than in the SAV-Oriented Scenario.

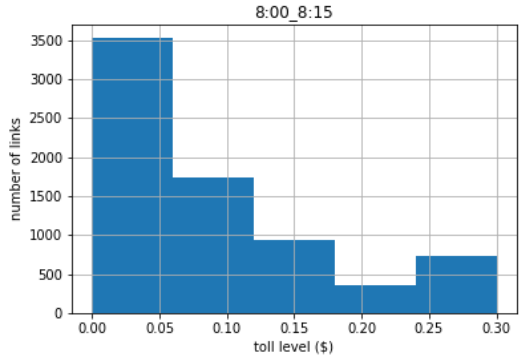

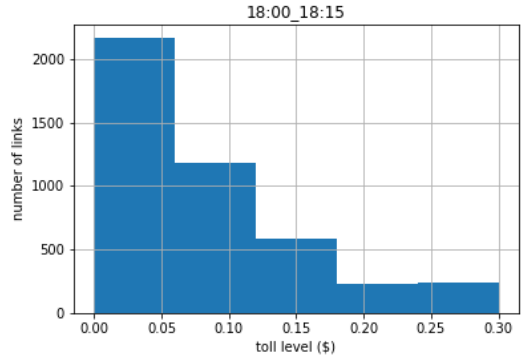

(a)

(b)

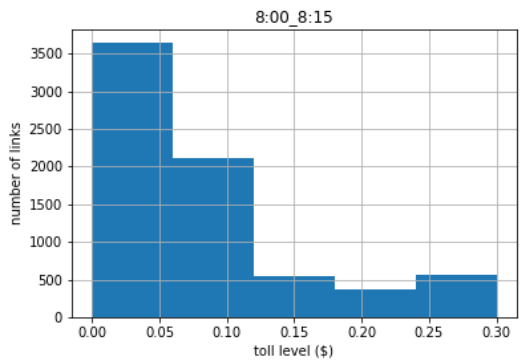

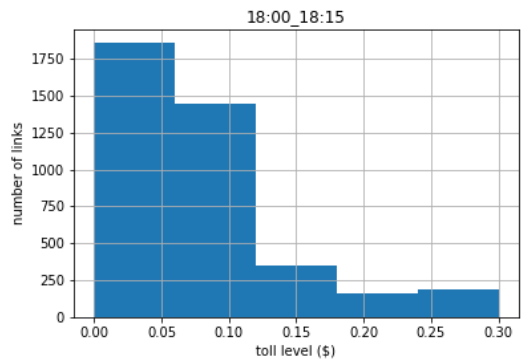

(c)

(d)

*Figure 8: Toll distribution during the morning peak and evening peak in the AV-Oriented Scenario (a-b) and SAV-Oriented Scenario (c-d)*

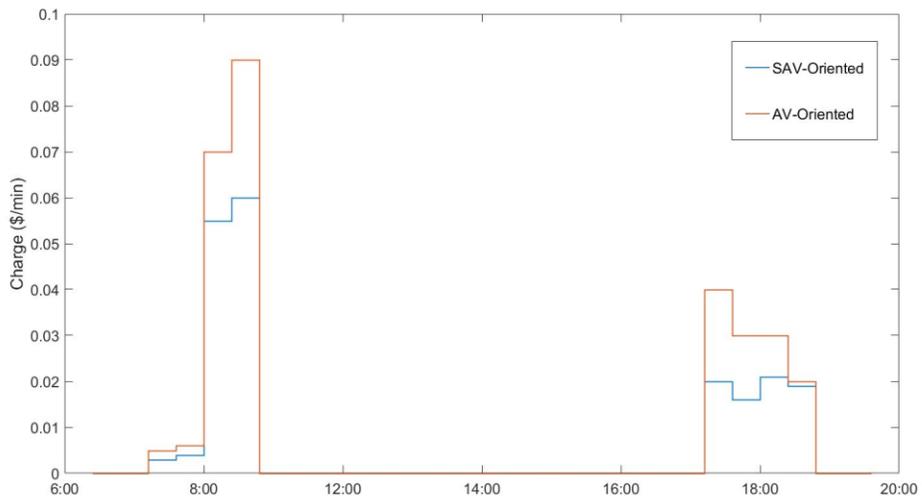

*Figure 9: Resulting tolls for the Travel Time-Congestion based scheme in the AV-Oriented and SAV-Oriented scenario*

# RESULTS AND IMPLICATIONS

The impacts derived from the different congestion pricing schemes in each scenario are discussed in this section. The evaluation of the schemes is carried out by means of a set of commonly used performance indicators such as mode shift, change of traffic delay and motorized trips. The analyses continue with a comparison of system welfare effects, followed by a discussion about the policy implications of the different schemes.

## Mode choice

All congestion-pricing strategies evaluated here succeed in reducing car, AV and SAV trips to different extents. Overall, PT and slow modes witness an increase in mode share (Tables 4 through 7), as expected – due to making road use more expensive.

Overall, the demand for conventional vehicles seems more elastic than the demand for SAVs and especially AVs, given the higher modal shift achieved for all the CP strategies. Because of their higher initial costs, car travelers are more incentivized than AV travelers to adopt PT or slow modes in the presence of tolls. SAV users face higher costs than AV users, so they are generally more responsive to tolls. For this reason, traditional CP strategies seem to be more effective in the Base Scenario (no AVs or SAVs) and the SAV-Oriented Scenario. The overall shift to active modes and public transit achieved by the Link-Based Scheme are comparable in the three scenarios only because of the higher charge in the AV-Oriented Scenario.

Among the traditional schemes, the Distance-based scheme generates larger changes in travelers' mode choice than the Link-based scheme in the Base Scenario. These results are in line with previous studies about distance-based schemes (Litman, 1999). Instead, the scenarios characterized by large presence of AVs and SAVs differ from each other in their modal shifts. While in the SAV-Oriented Scenario the two schemes have comparable effects, in the AV-Oriented Scenario the Link-based scheme reduces AV trips more than the Distance-based scheme does. This is an interesting outcome, since the two schemes are conceptually very different from one another and could have very different effects in terms of economic gains, distributional effects, and public acceptability. Distance-based strategies are usually much simpler for users to comprehend, but much less effective at combatting congestion, which is link and time of day specific

The MCP-based scheme induces less travel behavior changes than the ones achieved with the traditional Link-based scheme, since the average levels of charge are lower. For the same reason, the Travel Time-Congestion based scheme determines a lower reduction of private trips than the Distance-based scheme, particularly in the SAV-Oriented Scenario.

*Table 4: Modal share from the link-based scheme[3]*

|  | AV Oriented | SAV Oriented | Base (no SAVs-AVs) |
|---|---|---|---|
| **Car trips (%)** | 8.55 (-7.0) | 48.16 (-5.1) | 78.13 (-6.4) |
| **PT trips (%)** | 16.66 (360.2) | 14.87 (82.2) | 12.37 (68.3) |
| **Walk/bike trips (%)** | 6.61 (284.3) | 7.44 (67.9) | 9.48 (3.3) |
| **AV trips (%)** | 67.55 (-19.8) | 5.62 (-4.6) | 0.00 |
| **SAV trips (%)** | 0.61 (-48.3) | 23.9 (-22.3) | 0.00 |

---

[3] Relative changes of mode share from the no-toll scenario are reported in parentheses.

*Table 5: Modal share from the distance-based scheme[3]*

|  | AV Oriented | SAV Oriented | Base (no SAVs-AVs) |
|---|---|---|---|
| **Car trips (%)** | 8.53 (-7.2) | 47.51 (-6.4) | 69.06 (-17.3) |
| **PT trips (%)** | 4.93 (36.2) | 14.95 (83.2) | 10.07 (37.0) |
| **Walk/bike trips (%)** | 2.48 (44.2) | 7.12 (60.7) | 20.85 (127.1) |
| **AV trips (%)** | 83.25 (-1.3) | 5.82 (-1.2) | 0.00 |
| **SAV trips (%)** | 0.78 (-48.3) | 24.54 (-20.2) | 0.00 |

*Table 6: Modal share from the MCP-based scheme[3]*

|  | AV Oriented | SAV Oriented | Base (no SAVs-AVs) |
|---|---|---|---|
| **Car trips (%)** | 8.49 (-7.6) | 47.37 (-6.6) | - |
| **PT trips (%)** | 11.05 (205.2) | 17.26 (111.5) | - |
| **Walk/bike trips (%)** | 4.93 (186.6) | 8.38 (89.2) | - |
| **AV trips (%)** | 74.85 (-11.2) | 5.40 (-8.3) | - |
| **SAV trips (%)** | 0.66 (-44.1) | 21.56 (-29.9) | - |

*Table 7: Modal share from the Travel Time-Congestion based scheme[3]*

|  | AV Oriented | SAV Oriented | Base (no SAVs-AVs) |
|---|---|---|---|
| **Car trips (%)** | 8.69 (-5.4) | 48.89 (-3.6) | - |
| **PT trips (%)** | 5.22 (44.2) | 11.62 (42.4) | - |
| **Walk/bike trips (%)** | 2.92 (69.8) | 6.37 (43.8) | - |
| **AV trips (%)** | 82.35 (-2.3) | 5.84 (-0.8) | - |
| **SAV trips (%)** | 0.80 (-32.2) | 27.26 (-11.3) | - |

**Network performance**

Both traditional and advanced CP strategies induce a significant reduction of private trips traveled by AVs, SAVs and cars (Figure 10). Schemes with a distance dependent fee component do not necessarily achieve the highest VMT reduction. For example, the Link-based scheme induces higher VMT reductions than the traditional distance-based scheme in the AV-Oriented Scenario, and the MCP scheme does the same in SAV-Oriented Scenario as well. Vice versa, the Distance-based scheme seems to yield higher improvements in the Base Scenario. The Travel Time-Congestion based scheme has the lowest effect on travel demand in both the AV-Oriented and the SAV-Oriented Scenarios.

However, this is just one perspective to evaluate the effects of road pricing strategies, as the changes in terms of network daily travel delay show (Figure 11). The results vary significantly according to strategy and scenario. Interestingly, in scenarios characterized by presence of AVs and SAVs, CP strategies targeting the critical links (i.e., the Link-based schemes) generate higher delay reductions than the distance-based scheme. In the Base Scenario however, the Distance-Based scheme achieves a higher delay reduction (in line with VMT reductions). This result can be partially explained by the fact that long AV-SAV trips (that would be affected by higher distance-based charges) are less incentivized to switch to low-quality modes like PT. Advanced CP schemes seem to achieve equal or higher travel delay reductions than traditional CP schemes. For example, the MCP-based scheme outperforms the corresponding traditional link-based scheme with reductions higher by 2 to 5 percentage points depending on the scenario. The Travel Time-Congestion-based scheme determines comparable reductions of delays to the corresponding Distance-based scheme (but at lower modal shifts). Although changes in the mode shift are lower in the AV-Oriented scenario than in the SAV-Oriented scenario, the decrease of delay is similar. In this case, users seem to be more willing to reroute and reschedule their trips rather than switching to public transit or active modes.

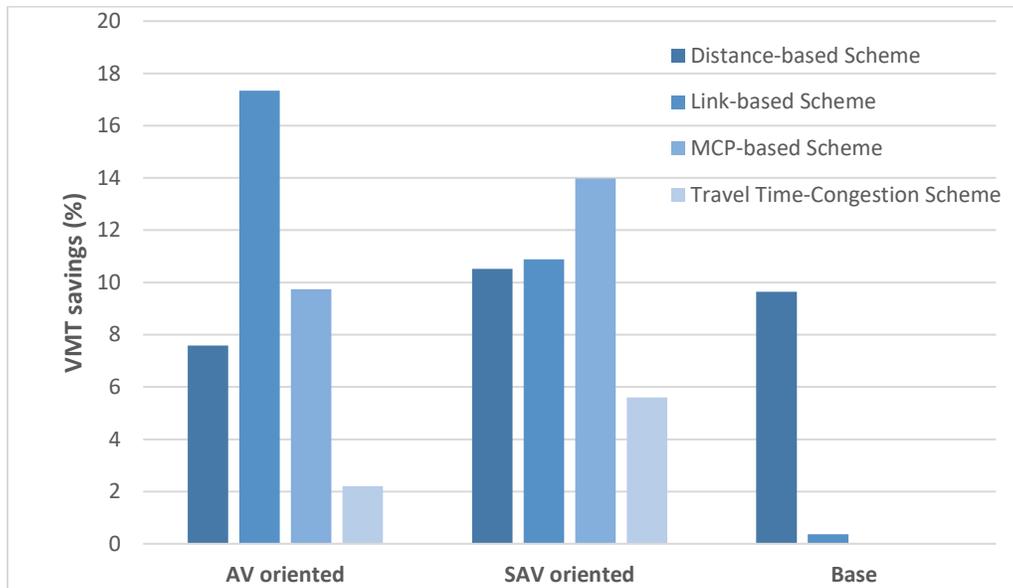

*Figure 10: Reduction of motorized trips for the different scenarios according to the congestion pricing scheme*

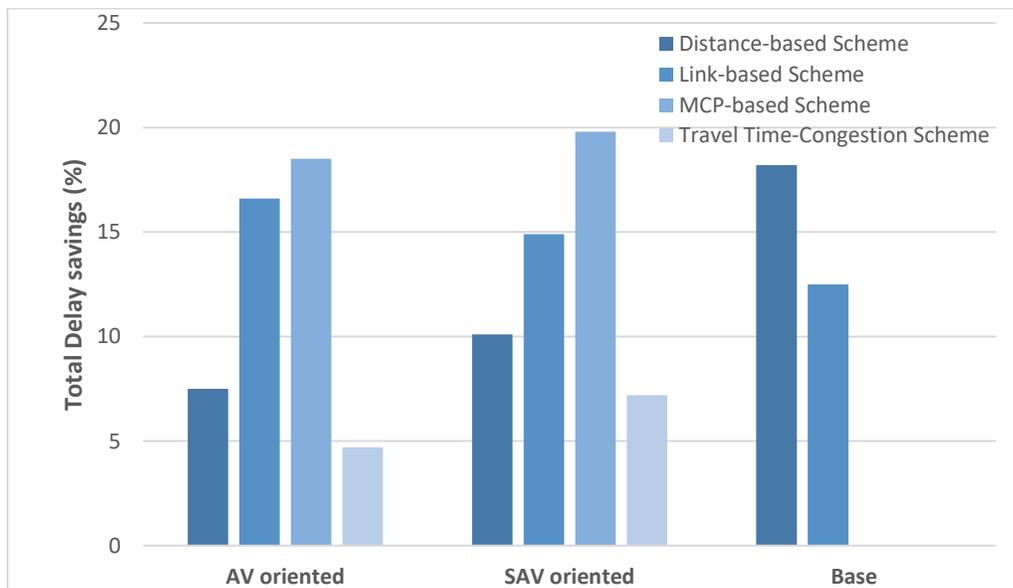

*Figure 11: Reduction of traffic delay for the different scenarios according to the congestion pricing scheme*

**Welfare changes**

Maximization of social welfare is important in evaluating transportation policy options. The social welfare change due to the introduction of pricing policies can be estimated using the "rule of the half," which approximates gains/losses of existing users (of facility by mode) and new users. The main drawbacks of this approach include the possibility of tolling's impacts on departure times and destinations, and users' heterogeneity (Arnott et al., 1990) in cases where economists do not reflect differences in users' willingness to pay for these different features of travel.

Multi-agent simulations like MATSim partly overcome these issues by allowing a "non-conventional" economic appraisal based on the agents' utilities as an economic performance indicator of the system. Although income differences are not considered here, it is possible to reflect different VTTS or marginal utilities of money by traveler.

In this study, the change in total welfare $\Delta\omega$ (across agents) between the original scenario (with no tolls) and each congestion pricing scenario is calculated as the sum of public revenues (first term in Eq. 13) and consumer surplus (second term in Eq. 13):

$$\Delta\omega = \tau + \sum_{j}^{J} \beta_m \cdot (V_j - V_j')$$

(13)

where $\tau$ is the sum of the collected toll revenues, $\beta_m$ is the negative marginal utility of monetary cost, and $V_j$ and $V_j'$ correspond to the total daily utility score for agent $j$ in the original scenario and in the congestion pricing scenario, respectively.

Table 8 summarizes the social welfare impacts of the different schemes for each scenario. The most effective strategies in terms of total welfare gains are the MCP-based scheme and the Travel Time-Congestion-based scheme. Both strategies perform better in the SAV-Oriented Scenario (assuming that the toll revenues could be fully reinvested). The higher levels of congestion of the AV-Oriented Scenario, the lower attractiveness of PT and active modes as compared to (private) autonomous transport, and the relatively long commute make CP strategies less inefficient. The Link-based scheme increases social welfare to a similar extent in all the scenarios. In contrast, the Distance-based scheme is found to improve total social welfare only in the SAV-Oriented Scenario and Base Scenario. Interestingly, the Travel Time-Congestion-based scheme yields the highest welfare gains by imposing relatively low fares. As expected, the advanced CP strategies seem to yield higher welfare improvements compared to the corresponding traditional ones. The MCP-based toll and Travel Time-Congestion respectively outperform the Link-based scheme and Distance-based scheme. Finally, social welfare changes in future scenarios characterized by different market developments of autonomous driving compare differently with the Base Scenario according to the typology of CP scheme. Link-based and Distance-based scheme in the SAV-Oriented Scenario have similar performance to the corresponding ones implemented in the Base Scenario. In the AV-Oriented Scenario, however, these two schemes are characterized by an opposite trend.

When the revenues are not considered, all of the CP strategies achieve a reduction in social welfare. In this case, the highest performance (in terms of the lowest reduction of consumer surplus) is achieved by the travel time-congestion-based scheme in both the AV-Oriented Scenario and SAV-Oriented Scenario. This is an important aspect to consider, since the ability to reinvest and the fraction of expendable revenues would determine whether a scheme is favorable, particularly from a public acceptance perspective.

*Table 8: Welfare changes for alternative CP schemes for each scenario*

| | AV-Oriented Scenario | SAV-Oriented Scenario | Base Scenario |
|---|---|---|---|
| **Original Scenario-total welfare (Million $/day)** | 7.264 | 5.340 | 14.242 |
| **Link-Based Scheme: consumer surplus change ( $ per capita per day)** | -3.10 | -1.07 | -0.26 |
| **Link-Based Scheme: welfare change with revenues ($ per capita per day)** | 0.08 | 0.07 | 0.10 |
| **Link-Based Scheme: Total welfare change (%)** | **0.48** | **0.47** | **0.69** |
| **Distance-Based Scheme: consumer surplus change ($ per capita per day)** | -1.70 | -1.39 | -1.00 |
| **Distance-Based Scheme: welfare change with revenues ($ per capita per day)** | -0.24 | 0.18 | 0.21 |
| **Distance-Based Scheme: Total welfare change (%)** | **-1.40** | **1.25** | **1.39** |
| **MCP-Based Scheme: consumer surplus change ($ per capita per day)** | -2.32 | -1.15 | - |
| **MCP-Based Scheme: welfare change with revenues ($ per capita per day)** | 0.33 | 0.43 | - |
| **MCP-Based Scheme: Total welfare change (%)** | **1.95** | **3.06** | - |
| **Travel Time-Congestion Based Scheme: consumer surplus change ($ per capita per day)** | -1.45 | -0.88 | - |
| **Travel Time-Congestion Scheme: welfare change with revenues ($ per capita per day)** | 0.44 | 0.63 | - |
| **Travel Time-Congestion Scheme: Total welfare change (%)** | **2.60** | **4.44** | - |

## Policy implications

All the investigated CP strategies significantly decrease traffic demand and reduce delays. Presumably, they will also reduce emissions, collisions, noise, and infrastructure damage.

However, only some CP strategies determined sizable gains in social welfare, and only after thoughtful compensation strategies are employed. Nevertheless, since gains from reduced emissions, noise and road damage were not considered in the analyses, the overall benefits from CP might be underestimated. In addition, Austin's relatively low transit service levels negatively affect the overall efficiency of CP strategies tested here, since drivers and SAV or AV users do not have reasonable alternative modes to consider (especially for longer trips, with distances above 5 km). Using revenues to improve delivery of public transit services is considered a progressive solution (from an equity perspective), at least for those with transit access (Ecola and Light, 2010). Such revenue use may result in net benefits for all strategies tested, and with far more winners than losers, at the level of individual travelers.

In general, CP strategies that target congestion at the link-level perform better in the Base and AV-Oriented Scenarios, while distance- and travel-time–based strategies appear more effective in the SAV-Oriented

Scenario. This suggests that travel demand management solutions may need to change as AV offerings evolve.

The shares of CP-based toll revenues that can be used to address social welfare costs are likely to be critical for distributional impacts and policy efficiency and, ultimately, public acceptability of CP policies. Indeed, in order to cope with increases in PT ridership due to CP pricing, some revenues may best be invested in PT system improvements. Other options include lump-sum transfers to residents, per-capita credits, and income-based discounts.

Finally, the overall welfare gains are relatively low (at most $0.63 per capita per day), even for the most beneficial (Travel Time-Congestion) strategy examined here. Particularly for the traditional strategies, the welfare gains might have been underestimated because the levels of charge have been derived by means of a relatively straightforward procedure. However, it is worth mentioning that these values are comparable to those obtained in similar studies (Safirova et al., 2004; Gulipalli and Kockelman, 2008) or real-world implementations (Eliasson and Mattsson, 2006). More precisely targeted tolls that vary according to traffic conditions within the peak periods yield greater welfare gains. These experiments suggest that it is best to focus on the longer-term GPS-based technologies for more advanced CP strategy implementation, across relatively large sections of our networks and more times of day. Under these circumstances, it becomes particularly important to quantify properly the investment and maintenance costs for different tolling strategies. For example, the employment of cheaper satellite and cellular technologies (compared to traditional tolling infrastructure) could lower the investment costs for advanced CP strategies and make them more attractive.

## CONCLUSION

AVs and SAVs will affect people's mobility and community's traffic conditions. In terms of congestion, it is not clear whether the benefits of increased accessibility and more efficient traffic flows will compensate for the cost of more trip-making and longer distances traveled. Congestion pricing schemes represent an opportunity to internalize the negative costs of traffic congestion. The evolving transportation landscape, eventually characterized by higher automation and connectivity, enables the implementation of relatively advanced CP strategies.

This study adopts an agent-based model to investigate the potential mobility, traffic and economic effects of different congestion pricing schemes in alternative future scenarios (one characterized by high adoption of AVs, the other by wide usage of SAVs) for the Austin, Texas metro area. In the two future scenarios analyzed, vehicle-miles traveled (VMT) and traffic delays rise due to mode shifts (away from traditional transit) and SAVs traveling empty.

From a traffic perspective, all the mobility schemes yield considerable reductions of congestion. While advanced CP schemes are not necessarily more effective than traditional ones in affecting travel demand and traffic, they bring higher economic gains. More importantly, the effects of different strategies vary depending on the scenario. The Distance-based based scheme seem more effective in the SAV-Oriented Scenario and in the Base Scenario, while the Link-based scheme performs better in the AV-Oriented Scenario. The MCP-based scheme and Travel Time-Congestion-based scheme perform better in the SAV-Oriented Scenario than in the AV-Oriented Scenario. In all the scenarios, the Travel Time-Congestion scheme yields the largest social welfare improvements.

The analysis of mobility scenarios by means of an agent-based model like MATSim allows a high level of realism since it is possible to explicitly model several factors concerning transportation demand and traffic. In the specific context of AVs-SAVs, the coexistence of different autonomous modes and cars is considered (in addition to public transit and walk/bike), as well as: the impacts of autonomous driving on increased capacity; the changes in travel costs and preferences, and the demand responsive mechanism of SAV services (with the phenomenon of empty trips).

The simulations performed in this study present some limitations. Some are being addressed in current research, while others are left for future research. As mentioned earlier, limited effort has been put into the calibration process to derive the scenarios' parameters, although all the chosen values are justifiable and the results realistic. The focus of this study is providing transparent and generalizable results that can be used as a benchmark for future studies. Making predictions for Austin's future mobility (involving AVs and SAVs), would require several other pieces of information such as land use, detailed AV ownership data, and gas price, and as such is beyond the scope of this research. It would be interesting to study how the results obtained in this study apply to other cities. Simulations could be further improved by explicitly modeling parking behavior and by directly accounting for parking costs, rather than using an ASC. In future studies of AV-SAV scenarios, it would be interesting to include the effects of automation on destination choice and the possibility of dropping activities in agents' plans as well.

In the specific field of travel demand management, additional studies can be performed to investigate the distributional effects of different CP schemes and possible compensation measures. The implementation of SAV-based dynamic ride-sharing services, their traffic impacts and their synergies with pricing strategies is another issue that is subject to ongoing research.

# APPENDIX I

*Table 9: Tests for different levels of charge for the Link-based scheme (best results in bold)*

| Link-Based Scheme | Levels of fare ($) | | |
|---|---|---|---|
| **Base Scenario** | **0.1** | **0.2** | **0.3** |
| Consumer Surplus Change ($ per capita per day) | **-0.26** | -0.37 | -0.36 |
| Consumer Surplus Change with revenues ($ per capita per day) | **0.11** | 0.02 | 0.03 |
| Total Welfare Change (%) | **0.7** | 0.1 | 0 |
| | | | |
| **AV-Oriented Scenario** | **0.1** | **0.2** | **0.3** |
| Consumer Surplus ($ per capita per day) | -2.36 | **-3.1** | -4.0 |
| Consumer Surplus Change with revenues ($ per capita per day) | -0.22 | **0.08** | -0.92 |
| Total Welfare Change (%) | -1.3 | **0.5** | -5.4 |
| | | | |
| **SAV-Oriented Scenario** | **0.1** | **0.2** | **0.3** |
| Consumer Surplus ($ per capita per day) | **-1.07** | -1.72 | -1.21 |
| Consumer Surplus Change with revenues ($ per capita per day) | **0.07** | -0.02 | -0.05 |
| Total Welfare Change (%) | **0.5** | -0.1 | -0.5 |

*Table 10: Tests for different levels of charge for the Distance-based scheme (best results in bold)*

| Distance-Based Scheme | Levels of fare ($ per mile) | | |
|---|---|---|---|
| **Base Scenario** | **0.1** | **0.2** | **0.3** |
| Consumer Surplus Change ($ per capita per day) | **-1.00** | -1.81 | -2.4 |
| Consumer Surplus Change with revenues ($ per capita per day) | **0.21** | -0.11 | -0.46 |
| Total Welfare Change (%) | **1.4** | -0.7 | -3.1 |
| | | | |
| **AV-Oriented Scenario** | **0.1** | **0.2** | **0.3** |
| Consumer Surplus ($ per capita per day) | **-1.70** | -3.28 | -4.61 |
| Consumer Surplus Change with revenues ($ per capita per day) | **-0.24** | -1.47 | -1.09 |
| Total Welfare Change (%) | **-1.4** | -8.7 | -6.4 |
| | | | |
| **SAV-Oriented Scenario** | **0.1** | **0.2** | **0.3** |
| Consumer Surplus ($ per capita per day) | **-1.39** | -2.60 | -3.62 |
| Consumer Surplus Change with revenues ($ per capita per day) | **0.17** | -0.08 | -0.64 |
| Total Welfare Change (%) | **1.2** | -0.8 | -4.5 |

**ACKNOWLEDGEMENTS**

The authors thank Michal Maciejewski and Amit Agarwal for fruitful discussions on the MATSim simulation, and Felipe Dias for support in the analyses. We are grateful to two anonymous referees for their constructive input. The study was partly funded by the Texas Department of Transportation under Project 0-6838, "Bringing Smart Transport to Texas".